\documentclass[conference,10pt]{IEEEtran}
\usepackage{amsmath,amsfonts}
\usepackage{algorithmic}
\usepackage{algorithm}
\usepackage{array}
\usepackage[caption=false,font=footnotesize,labelfont=rm,textfont=rm]{subfig}
\usepackage{textcomp}
\usepackage{stfloats}
\usepackage{url}
\usepackage{verbatim}
\usepackage{graphicx}
\usepackage{booktabs}
\usepackage{multirow}
\usepackage[nocompress,nospace]{cite}
 
\usepackage{siunitx}
\usepackage{import}
\usepackage{tikz}
\usepackage{pgfplots}

\usepackage[acronym,nonumberlist]{glossaries}
\usepackage{glossaries-prefix}
\usepackage{amssymb}
\usepackage{import}
\newacronym[shortplural=GMMs]{GMM}{GMM}{Gaussian mixture model}
\newacronym[shortplural=HMMs]{HMM}{HMM}{hidden Markov model}
\newacronym[shortplural=DNNs]{DNN}{DNN}{deep neural network}
\newacronym[shortplural=SVDs]{SVD}{SVD}{singular value decomposition}
\newacronym[
    prefixfirst={a\ },
    prefix={an\ }
]{MCTS}{MCTS}{Monte Carlo tree search}
\newacronym[prefixfirst={a\ },prefix={an\ }]{MDP}{MDP}{Markov decision process}
\newacronym{CMDP}{CMDP}{constrained Markov decision process}
\newacronym{RL}{RL}{reinforcement learning}
\newacronym[shortplural=DTs]{DT}{DT}{decision tree}
\newacronym{SMT}{SMT}{satisfiability modulo theories}
\newacronym{IL}{IL}{Imitation Learning}
\newacronym[shortplural=CNNs]{CNN}{CNN}{convolutional neural network}
\newacronym[shortplural=DQNs]{DQN}{DQN}{deep Q-network}
\newacronym{AI}{AI}{artificial intelligence}
\newacronym{PPO}{PPO}{proximal policy optimization}
\newacronym{UE}{UE}{user equipment}
\newacronym{BS}{BS}{base station}
\newacronym{MIMO}{MIMO}{multiple-input/multiple-output}
\newacronym{MISO}{MISO}{multiple-input/single-output}
\newacronym{RSS}{RSS}{received signal strength}
\newacronym{NR}{NR}{new radio}
\newacronym{3DPF}{3DPF}{3-dimensional peak finding}
\newacronym{CF}{Cf}{collaborative filtering}

\def\clap#1{\hbox to 0pt{\hss#1\hss}}

\usepackage[normalem]{ulem}

\def\ve#1{{\mathchoice{\mbox{\boldmath$\displaystyle #1$}}%
              {\mbox{\boldmath$\textstyle #1$}}%
              {\mbox{\boldmath$\scriptstyle #1$}}%
              {\mbox{\boldmath$\scriptscriptstyle #1$}}}}

\def\T{\mathsf{T}}

\def\defeq{\stackrel{\mbox{\tiny def}}{=}}

\definecolor{safeblue}{HTML}{0072B2}
\definecolor{safeorange}{HTML}{D55E00}
\definecolor{safebeige}{HTML}{E69F00}
\definecolor{safecyan}{HTML}{56B4E9}
\definecolor{safegreen}{HTML}{009E73}
\definecolor{safeyellow}{HTML}{F0E442} 
\definecolor{safeplum}{HTML}{CC79A7}
\definecolor{safepurple}{HTML}{332288}

\definecolor{mygray}{HTML}{687681}

\pgfplotsset{%
compat=1.17,
width=\linewidth,
yminorticks=true,
separate axis lines,
axis background/.style={fill=white},
xmajorgrids,
ymajorgrids,
grid style={loosely dotted, black},
tick style={color=black},
label style={font=\footnotesize\rmfamily},  
tick label style={font=\small},
legend style={font=\scriptsize\sffamily, cells={anchor=west}, legend plot pos=left, draw=mygray},
legend pos=south east}

\IEEEoverridecommandlockouts

\begin{document}

\title{Efficient Beam Search for Initial Access Using Collaborative Filtering}
\author{%
\thanks{This work was supported by the Fraunhofer Lighthouse project ``6G SENTINEL'', by the Bavarian Ministry for Economic Affairs, Infrastructure, Transport and Technology through the Center for Analytics-Data-Applications (ADA-Center) within the framework of ``BAYERN DIGITAL II'', and by the Federal Ministry of Education and Research of Germany in the programme of ``Souver\"an. Digital. Vernetzt.'' joint project 6G-RIC (16KISK020K).}
\IEEEauthorblockN{George Yammine\textsuperscript{\textsection}, Georgios Kontes\textsuperscript{\textsection}, Norbert Franke, Axel Plinge and Christopher Mutschler}\vspace{1mm}
\IEEEauthorblockA{Fraunhofer IIS, Fraunhofer Institute for Integrated Circuits IIS, Division Positioning and Networks, Nuremberg, Germany \\
\tt{\small{\{firstname.lastname\}@iis.fraunhofer.de}}}}
\maketitle
\begingroup\renewcommand\thefootnote{\textsection}
\footnotetext{Equal contribution.}
\endgroup

\begin{abstract}
Beamforming-capable antenna arrays overcome the high free-space path loss at higher carrier frequencies. However, the beams must be properly aligned to ensure that the highest power is radiated towards (and received by) the user equipment (UE).
While there are methods that improve upon an exhaustive search for optimal beams by some form of hierarchical search, they can be prone to return only locally optimal solutions with small beam gains. Other approaches address this problem by exploiting contextual information, e.g., the position of the UE or information from neighboring base stations (BS), but the burden of computing and communicating this additional information can be high. Methods based on machine learning so far suffer from the accompanying training, performance monitoring and deployment complexity that hinders their application at scale.

This paper proposes a novel method for solving the initial beam-discovery problem. It is scalable, and easy to tune and to implement. Our algorithm is based on a recommender system that associates groups (i.e., UEs) and preferences (i.e., beams from a codebook) based on a training data set. Whenever a new UE needs to be served our algorithm returns the best beams in this user cluster.
Our simulation results demonstrate the efficiency and robustness of our approach, not only in single BS setups but also in setups that require a coordination among several BSs. Our method consistently outperforms standard baseline algorithms in the given task.

\end{abstract}
\glsresetall


\section{Introduction}

The increased demand for higher data rates in mobile communications requires sophisticated methods and algorithms of the mobile communication network. In particular, 5G \gls{NR} allows to communicate over the millimeter-wave frequency spectrum. However, among other effects, the free-space path loss for such frequencies poses a huge challenge. Beamforming-capable antenna arrays with a large number of elements with large gain factors can be employed to overcome such issues. In practice, instead of feeding each antenna element with its individual radio-frequency frontend chain we prefer an \emph{analog beamforming network} where both the amplitude and phase of the signal are adjusted, before feeding each antenna element. This enables the antenna array to spatially point the main lobe of the resultant beam. One task in the communication system is then to properly select out of all possible beams, i.e., from a \emph{beam codebook}, one to serve the \gls{UE}, e.g., a mobile phone or similar device. This task is commonly referred to as \emph{beam management}. 

As next generation networks are envisioned to have larger codebooks and dense deployments~\cite{sentinel2021road}, the complexity of this vast ecosystem of hardware components and software services will be overwhelming, since beam management will be responsible for selecting the most suitable beam of a \gls{BS} for all \glspl{UE} in range. In this paper, we address the fundamental initial problem in a beam management process, the \emph{initial beam discovery}, which deals with the determination of an initial access beam to serve a \gls{UE} once it is enters the serving range of a \gls{BS}.
\begin{figure}[t]
\centering

\def\svgwidth{0.9\columnwidth}
\import{figs/}{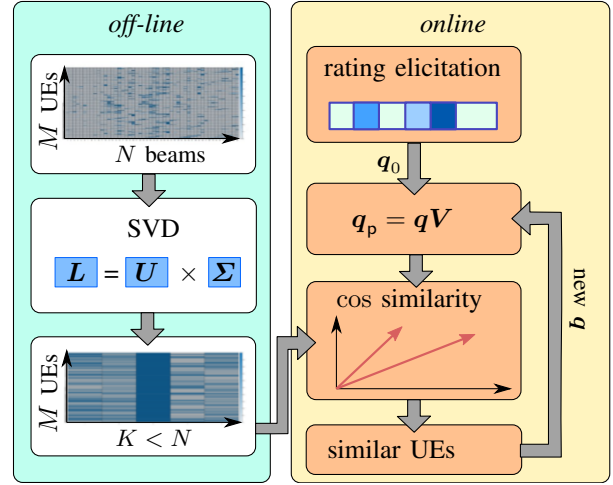}
\caption{The proposed initial beam-discovery framework based on the col\-la\-bor\-ative filtering approach. An SVD representation of the beams is calculated beforehand. For a new UE, an initial beam is selected by rating elicitation. This is then iteratively refined by choosing the best match via cosine similarity.}
\label{fig:cf}
\end{figure}

Even though the standard practice is to evaluate all of the beams~\cite{3gpp2020technical}, either via some form of exhaustive search~\cite{exhaustiveSearch2017} or hierarchical search \cite{desai2014initial}, significant research effort has been pursued towards predicting the optimal initial access beam with as less beam queries/probes as possible, to reduce measurement and reporting latency and network energy consumption~\cite{5gsurvey2022}. However, many such approaches rely on supervised learning using neural networks and suffer from scalability issues, due to their complex training and deployment.

This paper adapts ideas from the field of recommender systems to solve the initial beam-discovery problem, cf.\ Fig.~\ref{fig:cf}. The basic principle of such a system is to aggregate ``similar'' users, i.e., users with similar interests in movies for example, and recommend to them movies they have not yet seen, but were liked by other users in the group/cluster. We propose to re-formulate this problem by replacing the distinct users with the different \glspl{UE} and the movies items with the beams of the codebook, thus being able to identify \glspl{UE} that would be served best by the same beams. As recommender systems algorithms are quite mature, well-studied, and have proven to scale to commercial applications of enormous size, our proposed approach enjoys the same properties.

The rest of the paper is organized as follows. Sec.~\ref{sec:related} discusses related work before Sec.~\ref{sec:method} introduces our approach. Sec.~\ref{sec:numerical} describes our simulation environment and setup, and Sec.~\ref{sec:numerical_results} discusses the numerical results. Sec.~\ref{sec:conclusion} concludes our paper.

\section{Related Work}
\label{sec:related}

There has been a plethora of methods tackling the initial beam-discovery problem~\cite{giordani2018tutorial}, evaluated under the \emph{discovery time}, i.e., the time they need to find a good serving beam (measured in terms of how many beams are probed by the \gls{UE}) and the quality of the final selected beam, e.g., by measuring the \gls{RSS}. Naturally, there is a trade-off between the two; the more beams from the codebook are evaluated the higher the chance a better beam (or the overall best) in terms of quality of service is discovered.

The most basic approach here, proposed by 5G \gls{NR}~\cite{3gpp2020technical}, is exhaustive search, a brute-force search method that suggests that all possible beams are evaluated/measured by the \gls{UE}. Of course this method is guaranteed to always discover the best beam (except in the case where a codebook is also designed or adapted~\cite{heng2021learningArxiv}), but naturally requires a time- and energy-intensive measurement process from the \gls{UE} side. 

\emph{Hierarchical search} approaches first use wider beams that cover the entire service area and progressively evaluate narrower beams until eventually a narrow beam is selected to serve the \gls{UE}. While this is an efficient search method, the main drawback is that the initial wide beams have a lower beamforming gain and thus potentially not provide a high enough signal-to-noise ratio to properly detect the \gls{UE}~\cite{giordani2016comparative,aykin2020efficient}.

Other methods reduce the discovery time by utilizing some form of contextual information. For instance,~\cite{capone2015context} uses a rough estimate of the \gls{UE} position to locally search in a sub-set of the codebook for the best beam and~\cite{ismath2021deep} uses information from neighboring \glspl{BS} for faster beam discovery. However, this requires sophisticated network components (e.g., a positioning module) and/or advanced communication between \glspl{BS}.

Notably different, \gls{3DPF}~\cite{aykin2020efficient} partitions the beam codebook and searches via gradient descent. An initial beam at the spatial ``center'' of the codebook, as well as all the neighboring beams are evaluated and in case a neighboring beam is better, the process continues iteratively to the respective sub-matrix of the codebook. \gls{3DPF} is simple and efficient but prone to local minima, which can be avoided if the codebook matrix is divided in two or more sub-matrices and independent \gls{3DPF} searches are performed there.

In recent years, the success of deep learning led to the adoption of similar approaches for the initial beam-discovery problem. Hashemi et al.~\cite{hashemi2018efficient} suggested applying contextual bandit algorithms, but since this is an online algorithm the system performance can be low during training. Apart from this, simpler~\cite{sim2020deep,cousik2021deep} or more complicated~\cite{alrabeiah2020deep,lin2019bsnet,sohrabi2021deep} supervised and active (deep) learning approaches have been utilized both in research community and in standardization efforts~\cite{huawei2022tdoc,apple2022tdoc}. Despite their remarkable results, these approaches are not easily scalable. Training such networks requires a laborious hyper-parameter tuning process that necessitates expert knowledge, even though it can be partly automated. In addition, the automated training, deployment and re-training in case of performance degradation of neural networks is not trivial to solve at scale.

Our proposed approach manages to address all the above shortcomings, by achieving small detection time and high-quality beams, while being scalable and easy-to-deploy by design, facilitating only a single hyper-parameter to be tuned. 

\section{Proposed Method}\label{sec:method}

For our initial beam-discovery approach, we utilize and adapt the \emph{collaborative filtering (CF)} algorithm for \emph{recommender systems}~\cite{ekstrand2011collaborative}. If we take movie recommendations as an example, the core concept is the \emph{(user, movie, rating)} tuple, i.e., a user in the system has given a specific movie a subjective like/dislike rating (e.g., a real number in $\left[0, 5\right]$). The underlying assumption of the CF algorithm is that there is an fundamental notion of ``similarity'' between users in the system, thus users can be aggregated based on their preferences and movies that were liked by a group of users can be recommended to ``similar'' ones. The set of all tuples for all available user/movie ratings is stored in the \emph{rating matrix}.

For the work presented in this paper, we make the following association: \glspl{UE} are a similar concept to \textit{users}, the beams of a pre-defined codebook are the equivalent to \textit{movies}, and the measurements that indicate the service quality for a specific beam --- in our case the \gls{RSS} values scaled in $\left[0, 1\right]$ --- are essentially the ``ratings.'' 

Assuming that we have data collected from a set of \glspl{UE} $\mathcal{U} \defeq \{u_0, \ldots, u_{M-1} \}$ that request an initial access beam and this beam is determined after an exhaustive beam search~\cite{3gpp2020technical, exhaustiveSearch2017}, then, in the \emph{offline} phase, we can define the equivalent of the rating matrix of $M$ UEs and the \gls{RSS} measurements for the $N$ codebook beams as the matrix 
\begin{equation}
    \ve{Y} \defeq \begin{bmatrix}   
                    \mathrm{RSS}_{0,0} & \cdots &  \mathrm{RSS}_{0,N-1} \\
                    \vdots             &  \ddots &  \vdots \\
                    \mathrm{RSS}_{M-1,0} & \cdots & \mathrm{RSS}_{M-1,N-1}
                  \end{bmatrix} \;,
\end{equation}
where $\mathrm{RSS}_{i,n}$ is the measurement for user $i$ obtained for beam $n$, as shown in the left part of Fig.~\ref{fig:cf}. We then apply a \gls{SVD} on the data matrix and choose only the relevant subspace, i.e., $\ve{Y} = \ve{U} \ve{\varSigma} \ve{V}^\T$, where $\ve{U} \in \mathbb{R}^{M \times K}, \ve{\varSigma} \in \mathbb{R}^{K \times K}$ and $\ve{V} \in \mathbb{R}^{N \times K}$, with $K$ the reduced column dimension ($K < N$). Finally we define the UE/beam latent space as $\ve{L} \defeq \ve{U}\ve{\varSigma}$.

In the \emph{online} phase shown in the right part of Fig.~\ref{fig:cf}, a new UE has performed measurements for a set of $N_{\mathsf{init}}$ beams (with $N_{\mathsf{init}} \ll N$) which are stored in a vector $\ve{q} \in \mathbb{R}^{1 \times N}$. The task is using this initial beam ``rating'' to suggest beams from the codebook that would result in stronger RSS without needing to perform an exhaustive search. To achieve this, we project the sparse initial rating vector $\ve{q}$ in the same space, i.e., $\ve{q}_\mathsf{p} = \ve{q} \ve{V}$. Next, we measure the distance of the projected UE vector $\ve{q}_\mathsf{p}$ with all rows $\ve{\ell}_i$, $i \in \{0,\ldots,M-1 \}$, of the latent matrix $\ve{L}$ using the cosine distance
\begin{equation}
    d_{\mathsf{cos},i} = 1 - \frac{\ve{q}_\mathsf{p}^{} \ve{\ell}_i^\T }{\|\ve{q}_\mathsf{p} \| \| \ve{\ell}_i \|} \;,
\end{equation}
and determine the closest UEs, i.e., the set of most ``similar'' ones $\mathcal{X} \defeq \{ u_j \mid  \tilde{d}_{\mathsf{cos},j},\, j = 0,\,\ldots,\,X-1 \}$, where $\tilde{d}_{\mathsf{cos},j}$ is the cosine distance after sorting in ascending order. The \gls{RSS} values of each candidate beam in the codebook are estimated using the weighted distance of every member of the  similar users cluster $\mathcal{X}$ and a suggestion for the next beam to be probed is offered to the new UE. This process continues until a maximum budget for beam measurements is reached. \emph{Note that the only hyper-parameter to be tuned in our algorithm is the number of ``similar'' UEs to include for determining the next best beam}.

An important aspect in this approach is how to select the initial beams so that the similarity matching with existing UEs is more accurate, a problem called \emph{rating elicitation} for cold users in the recommender systems field. Here, several approaches have been proposed in the literature~\cite{elahi2011rating}, but in our work we select to always measure a specific pattern in the codebook. Even though more elaborate initial beam-probing methods exist~\cite{huawei2022tdoc}, our results indicate that a carefully selected pattern can lead to accurate results, as also indicated in 3GPP Rel.~18~\cite{huawei2022tdoc, apple2022tdoc}.

\section{Numerical Analysis Setup}
\label{sec:numerical}

\subsection{The Simulation Environment}

We used QuaDRiGa~\cite{quadriga1} to set up a simulation environment in \textsc{Matlab} with a geometry that is representative of a typical urban scenario: we place four base stations in the environment and let their panels point into a specific direction.

\begin{figure}[t]
\centering
\input{figs/paper_fig2.tex}
\caption{The beam codebook and indices used for each of the BSs: each row is an elevation angle and each column is an azimuth angle; initial probing pattern, i.e., on which beams to measure the \gls{RSS} initially, highlighted in red.} 
\label{fig:codebook}
\end{figure}

For the simulation we chose a carrier frequency of 28\,\si{GHz} a bandwidth of 100\,\si{MHz}, an OFDM modulation, and 64 active sub-carriers. On the transmitter side, we chose antenna-array panels with $8 \times 8$ elements (half-wavelength separated). We assume analog beamforming and each panel uses the same codebook with 60 beam directions. The main lobes of the beams point in steps of 10\textdegree\ from -45\textdegree\ to 40\textdegree\ in azimuth, and -50\textdegree\ to 0\textdegree\ in elevation. 
Fig.~\ref{fig:codebook} shows the indexing of the beams. The \gls{UE} is equipped with an omni-directional antenna and moves around the simulation environment.

\begin{figure}[t]
    \centering
    \includegraphics[width=0.72\linewidth]{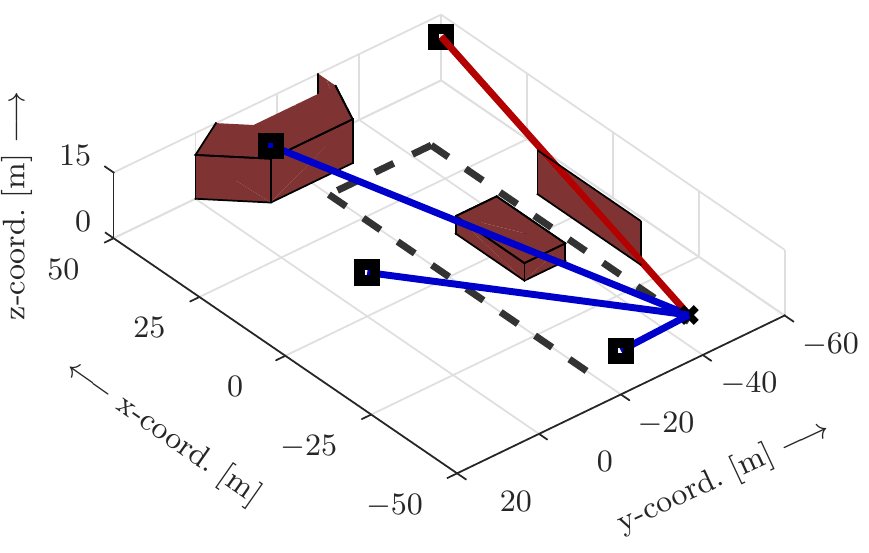}
    \caption{Visualization of the simulation environment. The UE (black cross) moves around the environment following the dashed path. At each step, the link between the BSs (black square) and the UE is checked for obstructions. Here, if a link is obstructed it is plotted in red and blue otherwise.}
    \label{fig:sim_env}
\end{figure}

We also add blockers that attenuate the received signal, and define them using different geometries: shape, size and location, see Fig.~\ref{fig:sim_env}. For our scenario, we assumed that each time the signal passes through one face of these objects, the signal strength is attenuated by 22\,\si{dB}. We chose our values based on the experimental data found in, e.g., \cite{attenuation}.

We define the \gls{RSS} indicator as the received power measured in the frequency domain. For our experiments, we assume a system without noise to compare the upper-bound performance of our approach and comparison benchmark. We also only assume a pure path-loss scenario, i.e., only line-of-sight (LOS) or obstructed LOS paths and no multipath signals.

To generate the channels we first define the environment layout (location of the base stations, the design and pointing of the antenna arrays, etc.) using QuaDRiGa. We then randomly defined a trajectory for the UE using 4 anchor points defining a ``U''-shaped track around the middle blocker. Next, for each spatial sampling point on this track, we project a ray from each BS to the UE. Using this ray, we count the intersections with blockers and attenuate the signal accordingly. Finally, we generate the frequency-domain channel coefficients per spatial sampling point and adjust them in case of attenuation.

\subsection{Baseline Algorithms for Comparison}
\label{sec:3dpf}

We evaluate the efficiency of our approach against two baseline algorithms. The first is the full beam sweep, i.e., the \gls{RSS} value of each beam in the codebook, shown in Fig.~\ref{fig:codebook}, is measured by the \gls{UE} and reported back to the \gls{BS}. This method is guaranteed to always discover the best initial beam --- at the cost of increased communication and energy requirements --- and we refer to it as the \emph{oracle}. The second is the 3DPF iterative algorithm~\cite{aykin2020efficient}. \gls{3DPF} starts from the \emph{approximate} spatial center of the codebook, i.e.,  beam 36 in our case, and then probes all neighboring beams (up, down, left, right). If the initial beam has the highest \gls{RSS} the search stops and this beam is returned, otherwise the sub-matrix with the best beam is selected and the search process repeats as before. As this algorithm may lead to sub-optimal solutions for problems with local optima, a parallel search can be performed in 2 or 3 sub-matrices and the best beam out of all is selected in the end. For our experiments we define three versions of the 3DPF algorithm that work on the codebook shown in Fig.~\ref{fig:codebook}: \textbf{3DPF/1} searches the entire 60-beam codebook, \textbf{3DPF/2} searches 2 sub-matrices in parallel (cols.\ 1--5 and 6--10), and \textbf{3DPF/3} searches 3 sub-matrices in parallel (cols.\ 1--3, 4--7 and 8--10).

For more information on the properties of the 3DPF algorithm, we refer the interested reader to~\cite{aykin2020efficient}.

\section{Numerical Results}
\label{sec:numerical_results}

\begin{table*}[b]
\caption{Available data sets used for the numerical simulations.\label{tab:dataset}}
\centering
\begin{tabular}{|c|c|c|c||c|c|c|c|}
\hline
\multicolumn{4}{|c||}{\textbf{Experimental setup}} & \multicolumn{4}{c|}{\textbf{Data properties}}\\
\hline
\textbf{Exp. ID} & \textbf{\# BS} & \textbf{full beam sweep train} & \textbf{full beam sweep CV}  & \textbf{\# training data} & \textbf{\# CV data} & \textbf{\# test data} & \textbf{sparsity (3DPF/1/2/3)}\\
\hline
1 (Sec.~\ref{sub:exp1}) & 1 & YES & YES & 1,962 / 9,810 & 4,360 & 17,440 & ---\\
\hline
2 (Sec.~\ref{sub:exp2}) & 4 & YES & YES & 1,962 / 9,810 & 4,360 & 17,440 & ---\\
\hline
3 (Sec.~\ref{sub:exp3}) & 1 & NO & YES & 1,962 / 9,810 & 4,360 & 17,440 & 79.4\% / 67.2\% / 57.5\%\\
\hline
4 (Sec.~\ref{sub:exp4}) & 1 & NO & NO & 17,876 & 21,800 & 17,440 & 79.5\% / 67.2\% / 57.5\%\\
\hline
\end{tabular}
\end{table*}

We simulate 1,000 UE trajectories with 218 time-steps each around the middle building, as exemplary shown (as a dashed line) in Fig.~\ref{fig:sim_env}. In each time-step, we perform a full beam sweep \emph{from all 4 BSs} to yield a plethora of available, fully labeled data. We pre-process the data by: i) scaling the \gls{RSS} values in $\left[0, 1\right]$; and ii) by adding a small positive value to \gls{RSS} measurements that are equal to zero in case of blocked UEs. The latter step is necessary for our algorithm to distinguish between low-quality beams and beams that have no \gls{RSS} measurements, as indicated by zero entries.

\begin{figure*}[t]
\centering
\subfloat[CF Performance BS1]{\begin{tikzpicture}
\begin{axis}[%
width=0.21\textwidth,
height=0.18\textwidth,
scale only axis,
xmin=12,
xmax=28,
ymin=0.5,
ymax=1,
xlabel={Avg.\ beam queries $\longrightarrow$},
ylabel={Avg.\ rel.\ performance $\longrightarrow$},
ytick={0.5, 0.6, 0.7, 0.8, 0.9, 1},
yticklabels={$0.5$, $0.6$, $0.7$, $0.8$, $0.9$, $1$},
xtick={12, 16, 20, 24, 28},
xticklabels={$12$, $16$, $20$, $24$, \clap{$28$}},
tick label style={font=\footnotesize},
]
\addplot [color=safeblue, line width=1.5pt, solid, mark=x]
 table[row sep=crcr] {%
12.3616 0.55024\\
19.7427 0.82945\\
25.4183 0.92483\\
 };%
\label{plot1}; 
\addplot [color=safeorange, line width=1.5pt, solid, mark=square]
 table[row sep=crcr] {%
12 0.74047\\
14 0.88897\\
16 0.93224\\
18 0.95349\\
20 0.9642\\
22 0.98019\\
24 0.98944\\
26 0.99464\\
 };%
\label{plot2};
\addplot [color=safebeige, line width=1.5pt, solid, mark=triangle]
 table[row sep=crcr] {%
12 0.73417\\
14 0.88201\\
16 0.94025\\
18 0.96823\\
20 0.97984\\
22 0.9883\\
24 0.99133\\
26 0.99347\\
 };%
\label{plot3};
\legend{}; 
\end{axis}
\end{tikzpicture}%
%
\hfill%
\label{fig:perf_BS1}}
\subfloat[CF Performance BS2]{\begin{tikzpicture}
\begin{axis}[%
width=0.21\textwidth,
height=0.18\textwidth,
scale only axis,
xmin=12,
xmax=28,
ymin=0.5,
ymax=1,
xlabel={Avg.\ beam queries $\longrightarrow$},
ylabel={},
ytick={0.5, 0.6, 0.7, 0.8, 0.9, 1},
yticklabels={},
xtick={12, 16, 20, 24, 28},
xticklabels={\clap{$12$}, $16$, $20$, $24$, \clap{$28$}},
tick label style={font=\footnotesize},
]
\addplot [color=safeblue, line width=1.5pt, solid,mark=x]
 table[row sep=crcr] {%
13.1487 0.56936\\
20.2243 0.84391\\
27.6981 0.92641\\
 };%
\addplot [color=safeorange, line width=1.5pt, solid,mark=square]
 table[row sep=crcr] {%
12 0.87393\\
14 0.92435\\
16 0.95419\\
18 0.96956\\
20 0.97913\\
22 0.98176\\
24 0.98337\\
26 0.98466\\
 };%
\addplot [color=safebeige, line width=1.5pt, solid,mark=triangle]
 table[row sep=crcr] {%
12 0.87784\\
14 0.92209\\
16 0.94631\\
18 0.96716\\
20 0.97811\\
22 0.98128\\
24 0.9838\\
26 0.98654\\
 };%
\end{axis}
\end{tikzpicture}%
%
\hfill%
\label{fig:perf_BS2}}
\subfloat[CF Performance BS3]{\begin{tikzpicture}
\begin{axis}[%
width=0.21\textwidth,
height=0.18\textwidth,
scale only axis,
xmin=12,
xmax=28,
ymin=0.5,
ymax=1,
xlabel={Avg.\ beam queries $\longrightarrow$},
ylabel={},
ytick={0.5, 0.6, 0.7, 0.8, 0.9, 1},
yticklabels={},
xtick={12, 16, 20, 24, 28},
xticklabels={\clap{$12$}, $16$, $20$, $24$, \clap{$28$}},
tick label style={font=\footnotesize},
]
\addplot [color=safeblue, line width=1.5pt, solid, mark=x]
 table[row sep=crcr] {%
13.4299 0.80778\\
21.7375 0.92864\\
28.5183 0.96143\\
 };%
\addplot [color=safeorange, line width=1.5pt, solid, mark=square]
 table[row sep=crcr] {%
12 0.91727\\
14 0.97358\\
16 0.99549\\
18 0.99812\\
20 0.99874\\
22 0.99886\\
24 0.9991\\
26 0.99987\\
 };%
\addplot [color=safebeige, line width=1.5pt, solid,mark=triangle]
 table[row sep=crcr] {%
12 0.91646\\
14 0.97096\\
16 0.99577\\
18 0.99784\\
20 0.99858\\
22 0.99953\\
24 0.99997\\
26 1\\
 };%
\end{axis}
\end{tikzpicture}%
%
\hfill%
\label{fig:perf_BS3}}
\subfloat[CF Performance BS4]{\begin{tikzpicture}
\begin{axis}[%
width=0.21\textwidth,
height=0.18\textwidth,
scale only axis,
xmin=12,
xmax=28,
ymin=0.5,
ymax=1,
xlabel={Avg.\ beam queries $\longrightarrow$},
ylabel={},
ytick={0.5, 0.6, 0.7, 0.8, 0.9, 1},
yticklabels={},
xtick={12, 16, 20, 24, 28},
xticklabels={\clap{$12$}, $16$, $20$, $24$, \clap{$28$}},
tick label style={font=\footnotesize},
]
\addplot [color=safeblue, line width=1.5pt, solid, mark=x]
 table[row sep=crcr] {%
11.8419 0.64852\\
17.3674 0.84007\\
23.0826 0.94258\\
 };%
\addplot [color=safeorange, line width=1.5pt, solid,mark=square]
 table[row sep=crcr] {%
12 0.74669\\
14 0.91542\\
16 0.97823\\
18 0.98889\\
20 0.99008\\
22 0.99347\\
24 0.99519\\
26 0.9996\\
 };%
\addplot [color=safebeige, line width=1.5pt, solid, mark=triangle]
 table[row sep=crcr] {%
12 0.76456\\
14 0.92371\\
16 0.9778\\
18 0.98648\\
20 0.9899\\
22 0.99357\\
24 0.99523\\
26 0.99968\\
 };%
\end{axis}
\end{tikzpicture}%
%
\label{fig:perf_BS4}}
\caption{Average relative performance vs.\ average number of beams queried for each single BS and all data available from exhaustive beam search. 3DPF search with 1,\,2,\,3 sub-sets (\ref{plot1}), CF with 1,962 UEs (\ref{plot2}) and CF with 9,810 UEs (\ref{plot3}).}
\label{fig:results_singleBS_fingerprint}
\end{figure*}
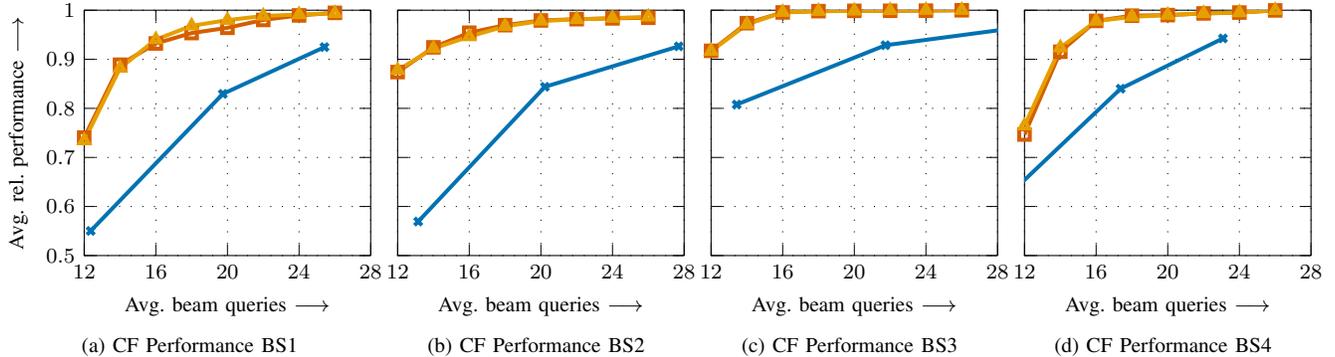

To evaluate our approach we conducted four numerical simulations under different assumptions. In the first simulation (Sec.~\ref{sub:exp1}), we assume that the BSs are completely independent. In the second simulation (Sec.~\ref{sub:exp2}), the BSs are able to coordinate and exchange data. In the third simulation (Sec.~\ref{sub:exp3}), we return to the single-BS scenario but training is performed using a sparse set obtained from 3DPF. In our final simulation (Sec.~\ref{sub:exp4}), we assume a sparse \emph{cross-validation (CV)} data set. Table~\ref{tab:dataset} lists the properties of the data sets.

As a final remark, in all experimental scenarios the final performance on the test set is evaluated as follows:
\begin{equation}
    \text{performance} \defeq \frac{1}{M_\mathsf{test}} \sum_{m=1} ^{M_\mathsf{test}} \frac{\mathrm{RSS}_\mathsf{algo}^m}{\mathrm{RSS}_\mathsf{oracle}^m} \;,
\end{equation}
where $M_\mathsf{test}$ is the size of the test set, $\mathrm{RSS}_\mathsf{algo}^m$ is the \gls{RSS} value obtained when using the beam selected by the algorithm to serve the UE, and $\mathrm{RSS}_\mathsf{oracle}^m$ is the \gls{RSS} value of the best beam. 

\subsection{Independent Base Stations}
\label{sub:exp1}

As a first experiment, we evaluate our algorithm for each BS independently, assuming that this is the only available serving BS, even in the case of blocked paths. After receiving the \gls{RSS} measurements for the initial probing beams (see Fig.~\ref{fig:codebook}) from the UE, the task is to suggest the most probable best beam.

We select the last 100 trajectories from the available data and use the first 20 of them (4,360 data points) as the CV set and the remaining 80 (17,440 data points) as the \emph{test set}. The test set contains data that is only used for the final beam prediction accuracy, hence not being used during training, while the CV set is used for tuning the single hyper-parameter of the algorithm (the number of ``similar'' UEs to be taken into account for suggesting the next best beam), thus avoiding over-fitting on the training data.

For evaluating the effect of the training data available in the final beam prediction performance, we generate two training sets of different sizes. The first contains 1,962 data points and it is generated by sampling every 100\textsuperscript{th} data point from the initial 900 available trajectories, while the second contains 9,810 data points after sampling every 20\textsuperscript{th} data point.

Fig.~\ref{fig:results_singleBS_fingerprint} shows the resulting performance of our algorithm relative to the oracle for all four BSs, when training and CV data are generated with the full beam sweep approach. In all cases our algorithm outperforms all versions of 3DPF. This is expected, as we utilize prior knowledge (training data) at the cost of data storage and the search process for ``similar'' UEs.

An interesting insight is that there is no significant performance gain with more training data. This is reasonable as despite the ``popular'' belief that more data yields better results, in reality data must be \emph{representative} for the task at hand. 
Here, even though one training data set is significantly smaller than the other, it correctly captures the spatial distribution of UEs in the test set and the actual application. 

\subsection{Coordinated Base Stations}
\label{sub:exp2}

Next, we evaluate our approach in a distributed BS coordination setting, as envisioned in future wireless network configurations~\cite{sentinel2021road}. 
Here, available data from full beam sweep for all \glspl{BS} are available, as in the first experimental setup, but the decision to be made is to determine which is the single best beam to serve the UE \emph{from all BSs}. In this setting, the train, CV, and test data contain the same UEs as before, but the respective rating matrices (see Fig.~\ref{fig:cf}) have 240 columns, including the codebooks of all 4 BSs.

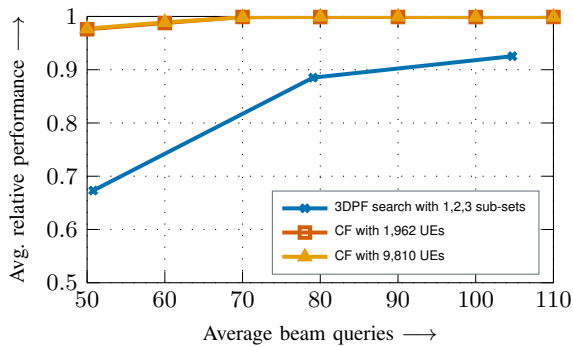
\begin{figure}[t]
\centering
\begin{tikzpicture}
\begin{axis}[%
width=0.7\columnwidth,
height=0.4\columnwidth,
scale only axis,
xmin=50,
xmax=110,
ymin=0.5,
ymax=1,
xlabel={Average beam queries $\longrightarrow$},
ylabel={Avg.\ relative performance $\longrightarrow$},
ytick={0.5, 0.6, 0.7, 0.8, 0.9, 1},
yticklabels={$0.5$, $0.6$, $0.7$, $0.8$, $0.9$, $1$},
xtick={50, 60, 70, 80, 90, 100, 110},
xticklabels={$50$, $60$, $70$, $80$,$90$,$100$, \clap{$110$}},
tick label style={font=\small},
]
\addplot [color=safeblue, line width=1.5pt, solid,mark=x]
 table[row sep=crcr] {%
50.7822 0.67308\\
79.0719 0.88516\\
104.7173 0.92537\\
 };%
\label{multi:plot1}; 
\addlegendentry{\tiny{3DPF search with 1,2,3 sub-sets}};
\addplot [color=safeorange, line width=1.5pt, solid,mark=square]
 table[row sep=crcr] {%
50 0.97596\\
60 0.98765\\
70 0.9991\\
80 0.9989\\
90 0.99902\\
100 0.99887\\
110 0.99906\\
 };%
\label{multi:plot2};
\addlegendentry{\tiny{CF with 1,962 UEs}};
\addplot [color=safebeige, line width=1.5pt, solid,mark=triangle]
 table[row sep=crcr] {%
50 0.97716\\
60 0.98898\\
70 0.99821\\
80 0.99925\\
90 0.99941\\
100 0.99866\\
110 0.99886\\
 };%
\label{multi:plot3};
\addlegendentry{\tiny{CF with 9,810 UEs}};
\end{axis}
\end{tikzpicture}%
\vspace{-1mm}
\caption{Avg.\ relative performance vs.\ avg.\ number of beams queried in case of multi-BS coordination (all data available from exhaustive beam search).}
\label{fig:perf_coop_multiBS_fingerprint}
\end{figure}

The same observations can be made for the multi-BS coordination case, with the resulting performance shown in Fig.~\ref{fig:perf_coop_multiBS_fingerprint}. Our CF algorithm again outperforms all versions of \gls{3DPF}. This is expected since \gls{3DPF} facilitates \emph{independent} smart search processes in each BS, while our method implicitly enables collaboration between the \gls{BS}. Again, the performance of the algorithm using both small and large training data sets is very similar, indicating the necessity for carefully selected training and CV data. After the initial beam pattern (Fig.~\ref{fig:codebook}) is evaluated in all \glspl{BS}, the next possible beams suggested depend on the \gls{RSS} measurements of the ``similar'' users in the training data. So if the \gls{UE} is close to one \glspl{BS}, similar \glspl{UE} would have \gls{RSS} patterns that indicate that beams from that \gls{BS} should be tried out next. Instead, if a \gls{UE} is between two \glspl{BS}, the beam probing strategy is evaluating beams from both \glspl{BS}, but without probing any beams from the other two. In other words, after the initial beam query, our proposed algorithm has a rough idea of strong \glspl{BS} to serve the \gls{UE} (based on the \gls{RSS} values of ``similar'' \glspl{UE}) and automatically allocates the remaining beam query budget there. 

\subsection{Training Using Sparse Training Data Set}
\label{sub:exp3}

\begin{figure}[t]
    \centering
    \begin{tikzpicture}
\begin{axis}[%
width=0.7\columnwidth,
height=0.4\columnwidth,
scale only axis,
xmin=12,
xmax=26,
ymin=0.5,
ymax=1,
xlabel={Average beam queries $\longrightarrow$},
ylabel={Avg.\ relative performance $\longrightarrow$},
ytick={0.5, 0.6, 0.7, 0.8, 0.9, 1},
xtick={12, 14, 16, 18, 20, 22, 24, 26},
xticklabels={$12$, $14$, $16$, $18$, $20$,$22$,$24$,\clap{$26$}},
tick label style={font=\small},
]
\addplot [color=safeblue, line width=1.5pt, solid,mark=x]
 table[row sep=crcr] {%
12.3616 0.55024\\
19.7427 0.82945\\
25.4183 0.92483\\
 };%
\label{cv:plot1}; 
\addlegendentry{\tiny{3DPF search with 1,2,3 sub-sets}};
\addplot [color=safeorange, line width=1.5pt, solid,mark=square]
 table[row sep=crcr] {%
12 0.7688\\
14 0.83589\\
16 0.89183\\
18 0.90621\\
20 0.91167\\
22 0.92125\\
24 0.92922\\
26 0.93679\\
 };%
\label{cv:plot2};
\addlegendentry{\tiny{CF with 3DPF/1 and 1,962 UEs}};
\addplot [color=safebeige, line width=1.5pt, solid,mark=triangle]
 table[row sep=crcr] {%
12 0.77748\\
14 0.8447\\
16 0.89594\\
18 0.9078\\
20 0.91213\\
22 0.91402\\
24 0.93079\\
26 0.938\\
 };%
\label{cv:plot3};
\addlegendentry{\tiny{CF with 3DPF/2 and 1,962 UEs}};
\addplot [color=safeplum, line width=1.5pt, solid,mark=diamond]
 table[row sep=crcr] {%
12 0.79956\\
14 0.86227\\
16 0.91892\\
18 0.94334\\
20 0.95642\\
22 0.96047\\
24 0.96507\\
26 0.96563\\
 };%
\addlegendentry{\tiny{CF with 3DPF/3 and 1,962 UEs}};
\addplot [color=safegreen, line width=1.5pt, solid,mark=o]
 table[row sep=crcr] {%
12 0.75466\\
14 0.84144\\
16 0.8926\\
18 0.89389\\
20 0.91311\\
22 0.91558\\
24 0.91927\\
26 0.927\\
 };%
\addlegendentry{\tiny{CF with 3DPF/1 and 9,810 UEs}};
\addplot [color=safecyan, line width=1.5pt, solid,mark=asterisk]
 table[row sep=crcr] {%
12 0.77381\\
14 0.8481\\
16 0.89652\\
18 0.90822\\
20 0.91275\\
22 0.91452\\
24 0.92909\\
26 0.93541\\
 };%
\addlegendentry{\tiny{CF with 3DPF/2 and 9,810 UEs}};
\addplot [color=safepurple, line width=1.5pt, solid,mark=pentagon]
 table[row sep=crcr] {%
12 0.79106\\
14 0.86436\\
16 0.92794\\
18 0.94417\\
20 0.95702\\
22 0.95957\\
24 0.96575\\
26 0.96592\\
 };%
\addlegendentry{\tiny{CF with 3DPF/3 and 9,810 UEs}};
\end{axis}
\end{tikzpicture}%
    \vspace{-1mm}
    \caption{Avg.\ relative performance vs.\ avg.\ number of beams queried for BS1 with sparse training data and full CV data set (from exhaustive beam search).}
    \label{fig:perf_sparse_fingerprint_BS1}
\end{figure}
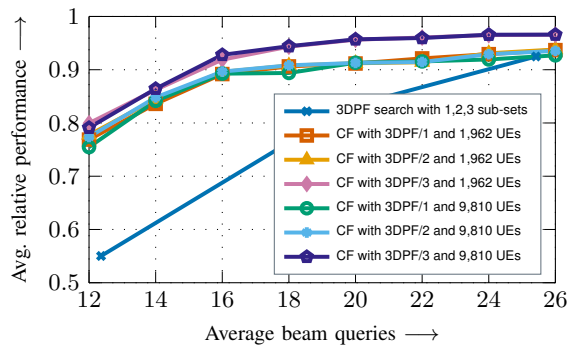

For our third experiment, we return our focus on single-BS settings. We have the same available data for each BS as in the first experimental setup, but we assume that the training data is not generated following the exhaustive beam search strategy, but a smart search algorithm or an already-trained machine learning model probing a smaller number of beams is in place. Here, the three different versions of 3DPF algorithm described in Sec.~\ref{sec:3dpf} are used to generate the data. Of course, in this case there is no guarantee that the best possible beam for each UE is included in the available training data. We still assume that we have a fully labeled CV set for hyper-parameter tuning.

In Fig.~\ref{fig:perf_sparse_fingerprint_BS1}, we again see that the performance of both small data sets is similar, but there is an important difference: for data generated with 3DPF/1 and 3DPF/2 (see Sec.~\ref{sec:3dpf}), the results for a beam budget over 16 beams are not as good as when full data is available (Fig.~\ref{fig:perf_BS1}). With less sparse data though, using 3DPF/3 for training data generation, there is a clear performance boost. This is a strong indication that our algorithm benefits from more complete and rich training data, as this additional information leads to more accurate determination of ``similar'' UEs. 

\subsection{Training Using Sparse Training And Cross Validation Data Sets}
\label{sub:exp4}

We further evaluate the case where both the training and CV data sets are sparse, i.e., generated by a smart search or a machine learning model making beam predictions. A preliminary study indicated that a small sparse CV set was not representative enough for successful hyper-parameter selection, so we opted for larger training and CV sets. We keep the same last 80 trajectories as a test set as before, but we use the first 820 trajectories for training (sampled every $10^{\text{th}}$ data point) and the following complete 100 trajectories for the CV data, resulting in 17,876 and 21,800 data points, respectively.

\begin{figure}[t]
    \centering
    \begin{tikzpicture}
\begin{axis}[%
width=0.7\columnwidth,
height=0.4\columnwidth,
scale only axis,
xmin=12,
xmax=26,
ymin=0.5,
ymax=1,
xlabel={Average beam queries $\longrightarrow$},
ylabel={Avg.\ relative performance $\longrightarrow$},
ytick={0.5, 0.6, 0.7, 0.8, 0.9, 1},
xtick={12, 14, 16, 18, 20, 22, 24, 26},
xticklabels={$12$, $14$, $16$, $18$, $20$,$22$,$24$,\clap{$26$}},
tick label style={font=\small},
]
\addplot [color=safeblue, line width=1.5pt, solid,mark=x]
 table[row sep=crcr] {%
12.3616 0.55024\\
19.7427 0.82945\\
25.4183 0.92483\\
 };%
\addlegendentry{\tiny{3DPF search with 1,2,3 sub-sets}};
\addplot [color=safeorange, line width=1.5pt, solid,mark=square]
 table[row sep=crcr] {%
12 0.7482\\
14 0.84152\\
16 0.89566\\
18 0.90881\\
20 0.91237\\
22 0.91536\\
24 0.91982\\
26 0.92134\\
 };%
\addlegendentry{\tiny{CF with 3DPF/1 and 17,876 UEs}};
\addplot [color=safebeige, line width=1.5pt, solid,mark=triangle]
 table[row sep=crcr] {%
12 0.77594\\
14 0.84252\\
16 0.89058\\
18 0.90127\\
20 0.90906\\
22 0.91141\\
24 0.91963\\
26 0.92041\\
 };%
\addlegendentry{\tiny{CF with 3DPF/2 and 17,876 UEs}};
\addplot [color=safeplum, line width=1.5pt, solid,mark=diamond]
 table[row sep=crcr] {%
12 0.78915\\
14 0.86198\\
16 0.91853\\
18 0.94571\\
20 0.93876\\
22 0.95375\\
24 0.95461\\
26 0.95558\\
 };%
\addlegendentry{\tiny{CF with 3DPF/3 and 17,876 UEs}};
\end{axis}
\end{tikzpicture}%
    \vspace{-1mm}
    \caption{Avg.\ relative performance vs.\ avg.\ number of beams queried for BS1 with sparse training and CV data.}
    \label{fig:perf_sparse_cv_BS1}
\end{figure}
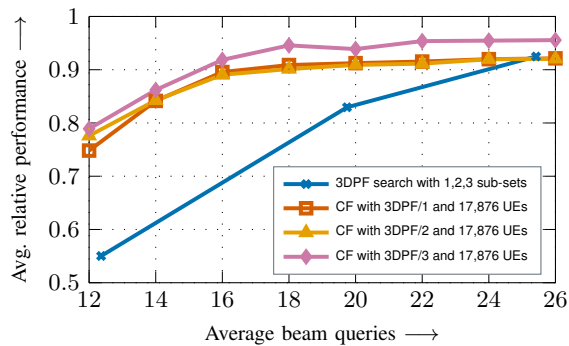

Fig.~\ref{fig:perf_sparse_cv_BS1} tells a similar story. For a beam budget over 16 beams, the performance of CF algorithm with more sparse data (generated with 3DPF/1/2) is reduced compared to the case where full data sets are available (Fig.~\ref{fig:perf_BS1}), while for data generated with 3DPF/3, performance is clearly increased. As both the training and CV data sets are now significantly larger compared to the previous studies, we once again verify that higher information gain in more representative data can be more important than the available data size. 

Despite the small performance drop from utilizing fully-labeled data sets to sparse data settings, we have to emphasize that our approach still outperforms all versions of 3DPF. This is important especially in the sparse setting, since the available training (and CV in the second case) data here are generated using this very algorithm, meaning we achieve higher performance compared to the algorithm used to generate the same data we utilize. This should not come as a surprise, as we extract more information from the available training data at the cost of data storage and search for ``similar'' UEs.

\section{Conclusion}\label{sec:conclusion}
\glsresetall%

We proposed a new, low-complexity approach based on recommender system theory for the initial beam-access/-discovery problem. We proved its efficiency in discovering the best initial beam in a challenging scenario with low beam-sweeping overhead, and showed, that it provides robust solutions even in settings with missing measurement data, while it can seamlessly scale, by design, to address complex settings that facilitate large codebooks and multi-BS coordination.

Future work integrates our algorithm into a learning-based beam management solution that addresses initial beam discovery, beam tracking, and hand-over tasks holistically.

\bibliographystyle{IEEEtranM} 
\bibliography{IEEEfull,conffull,references} 

\end{document}